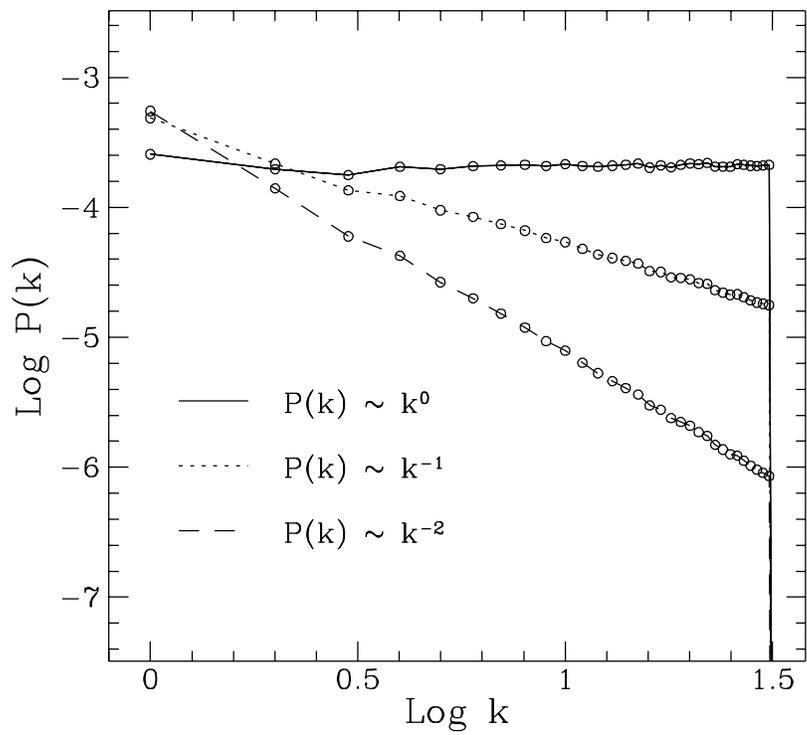

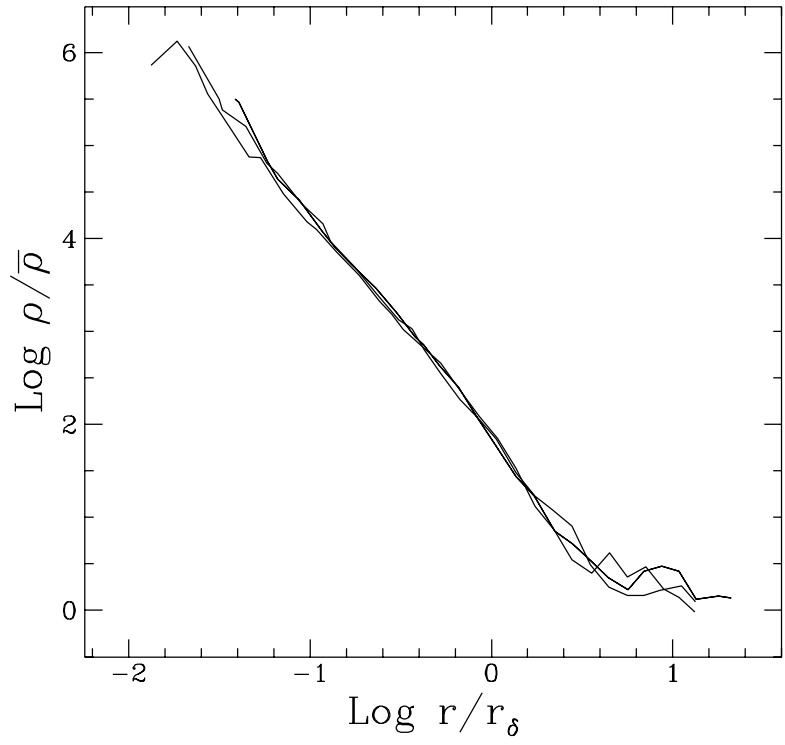

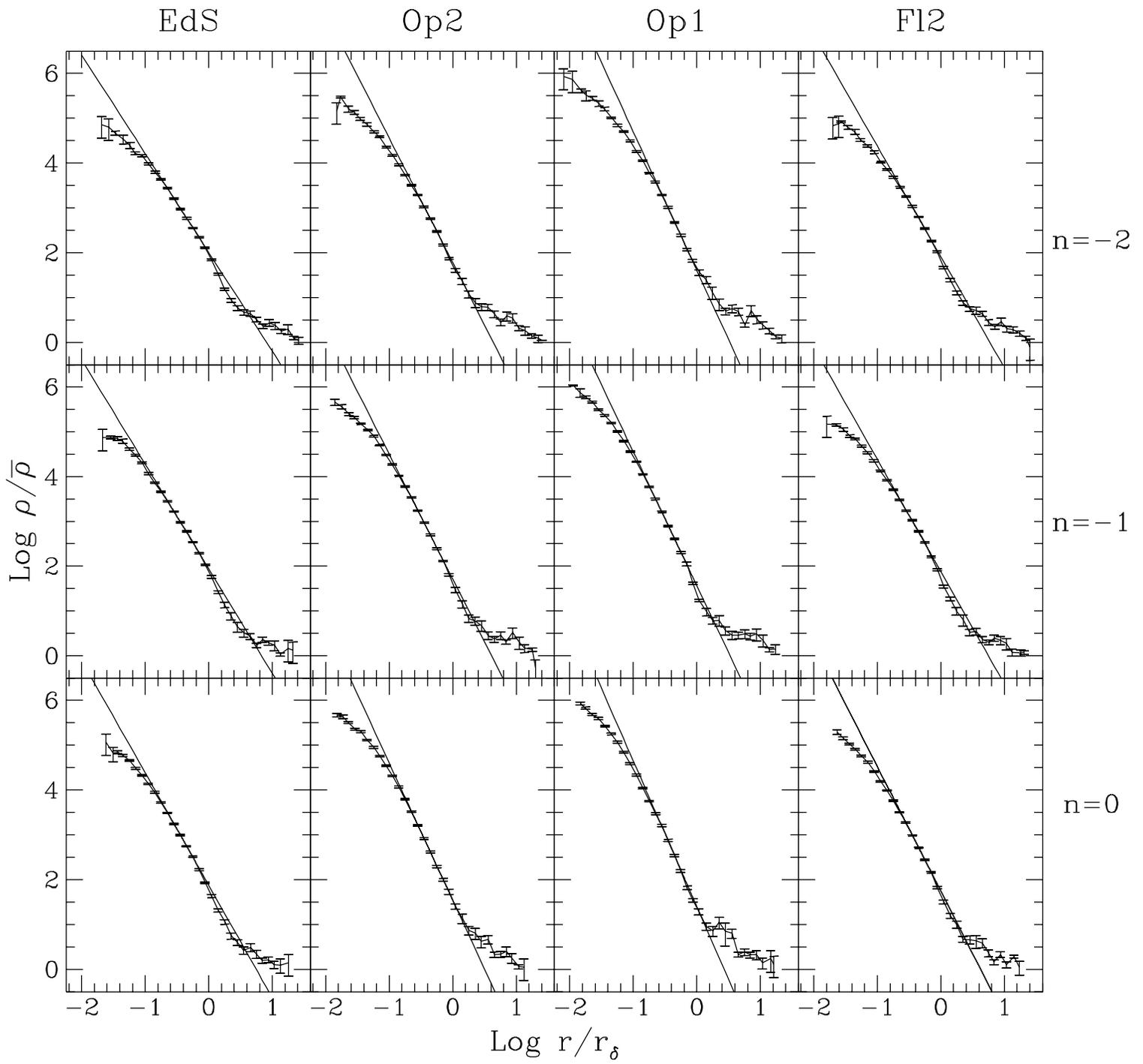

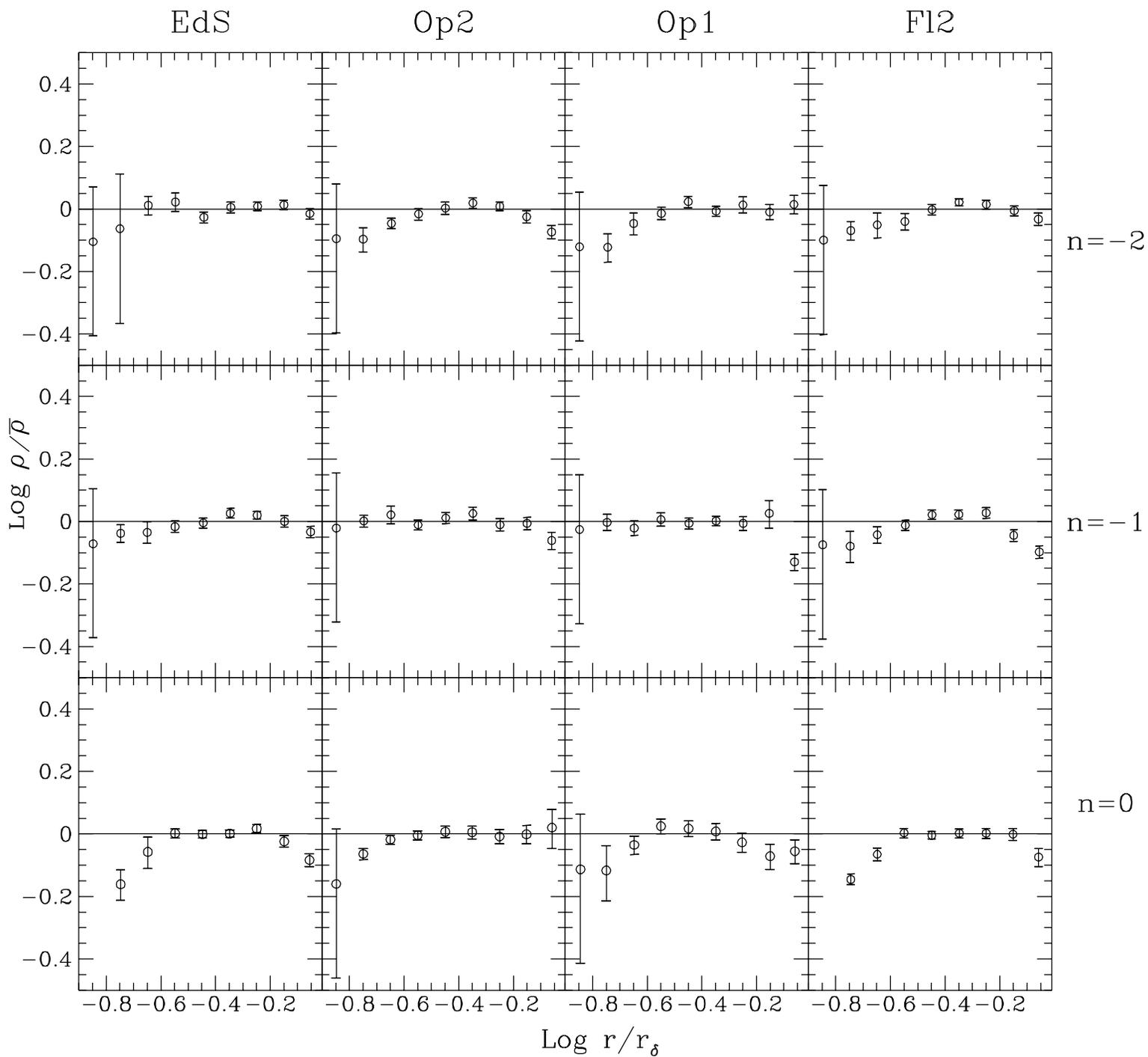

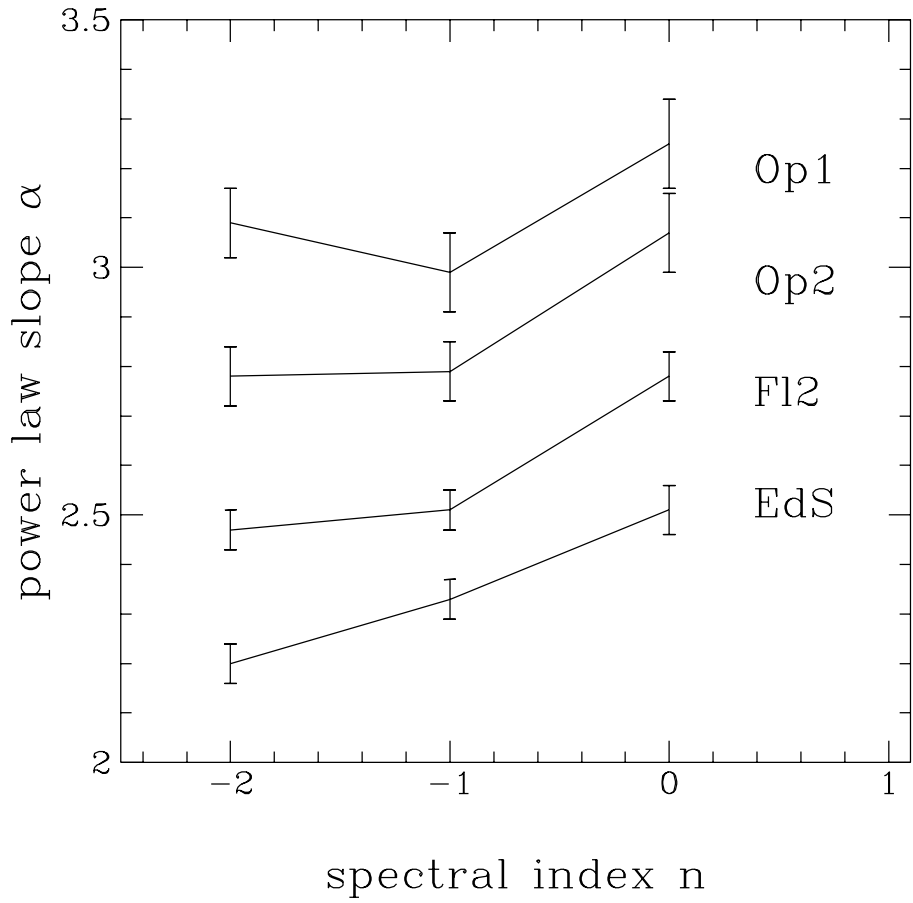

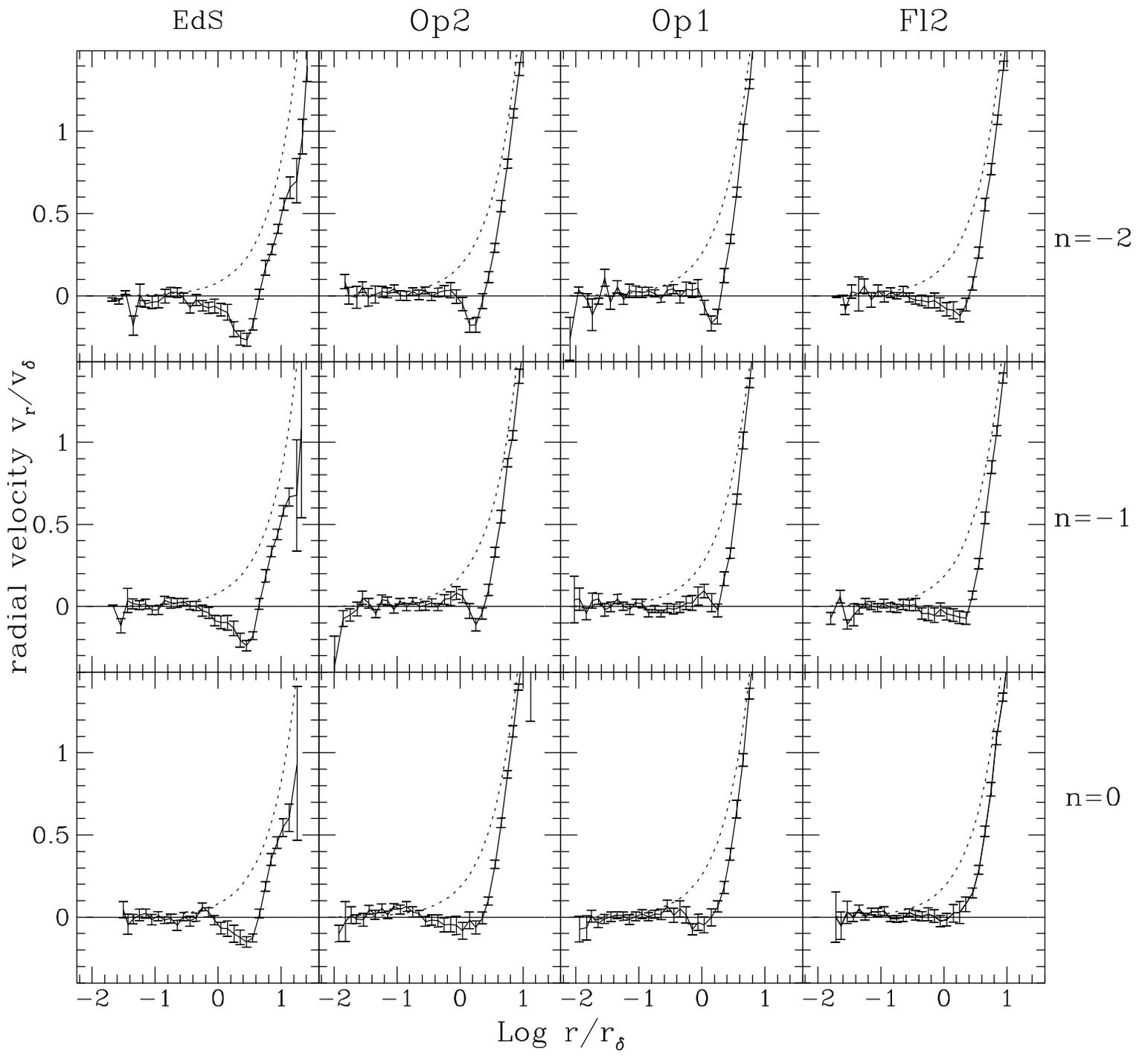

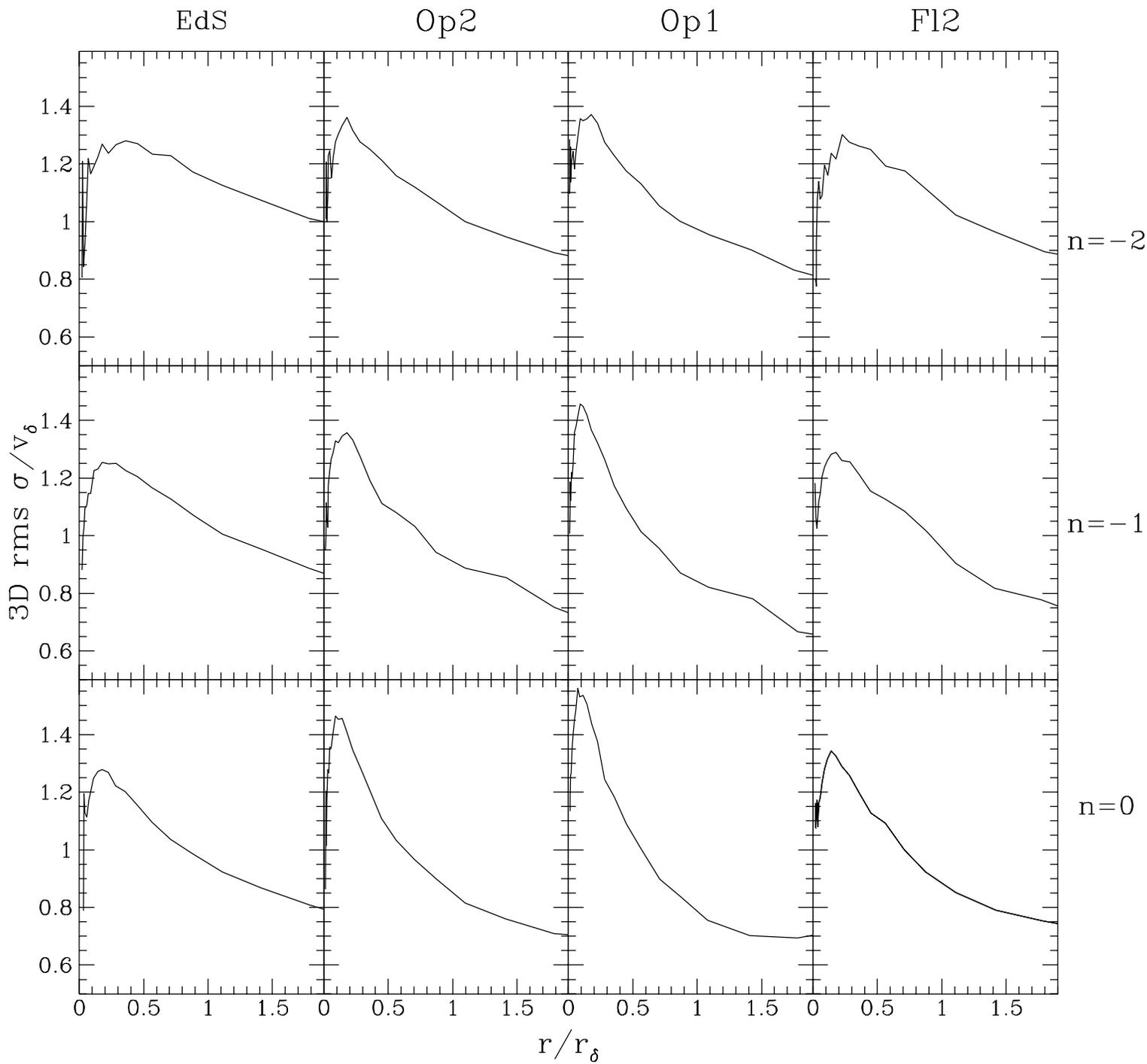

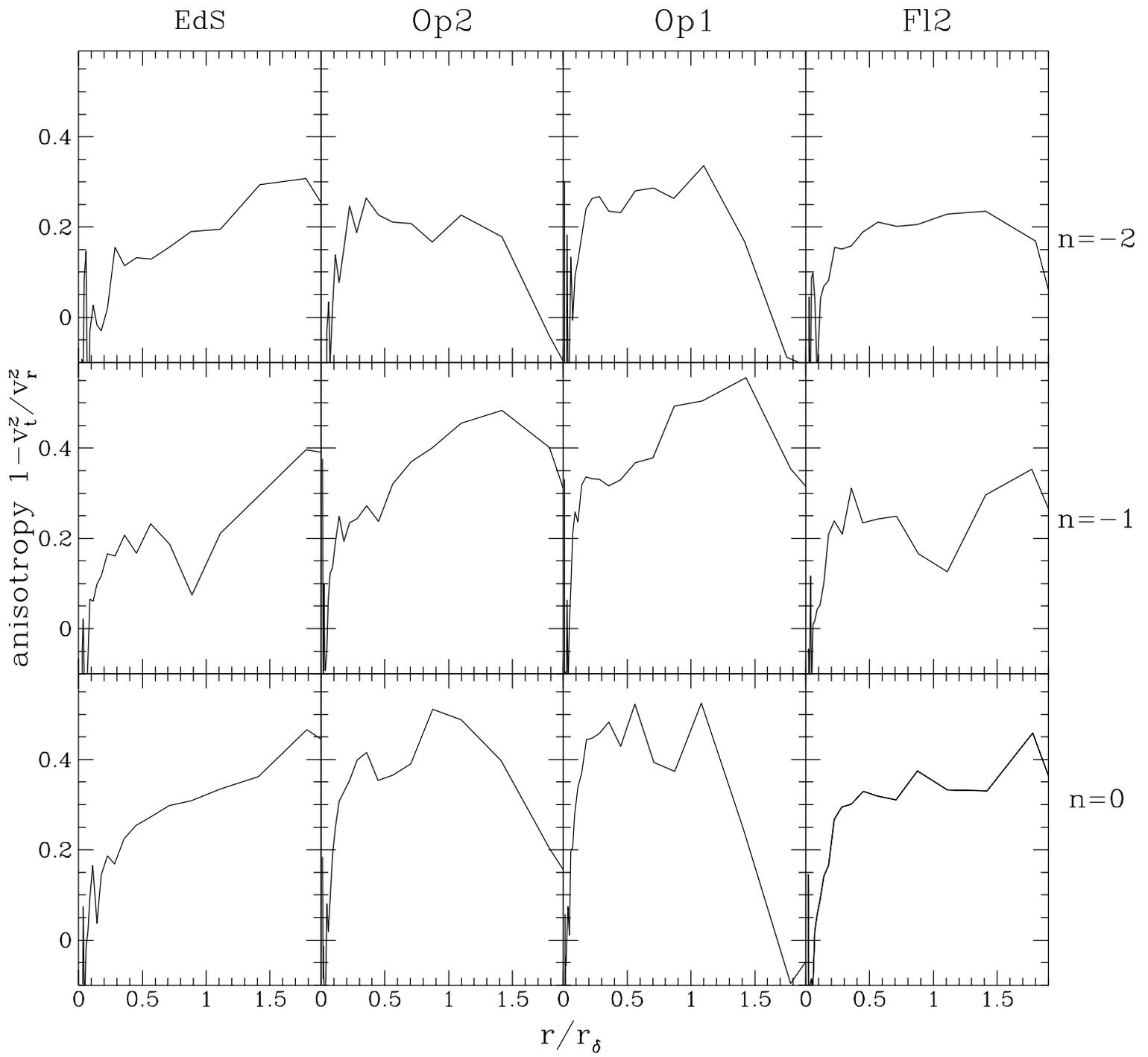

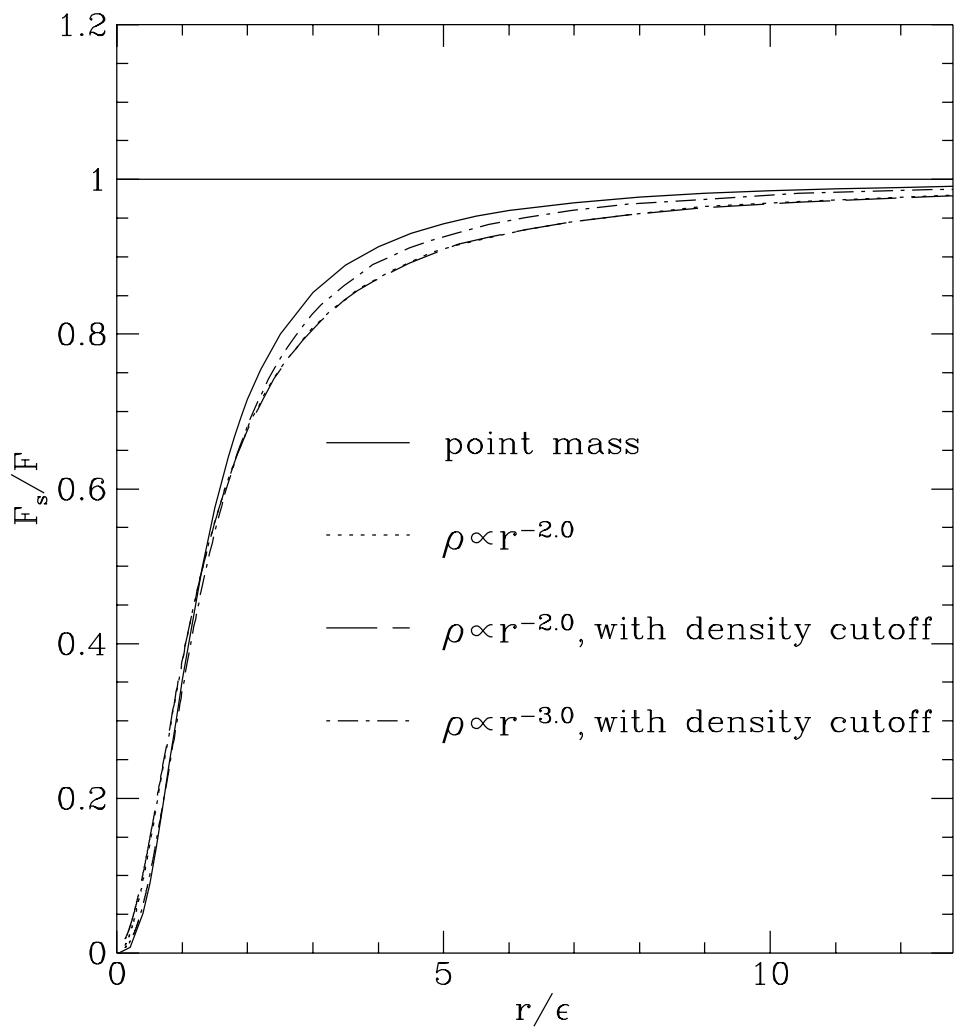

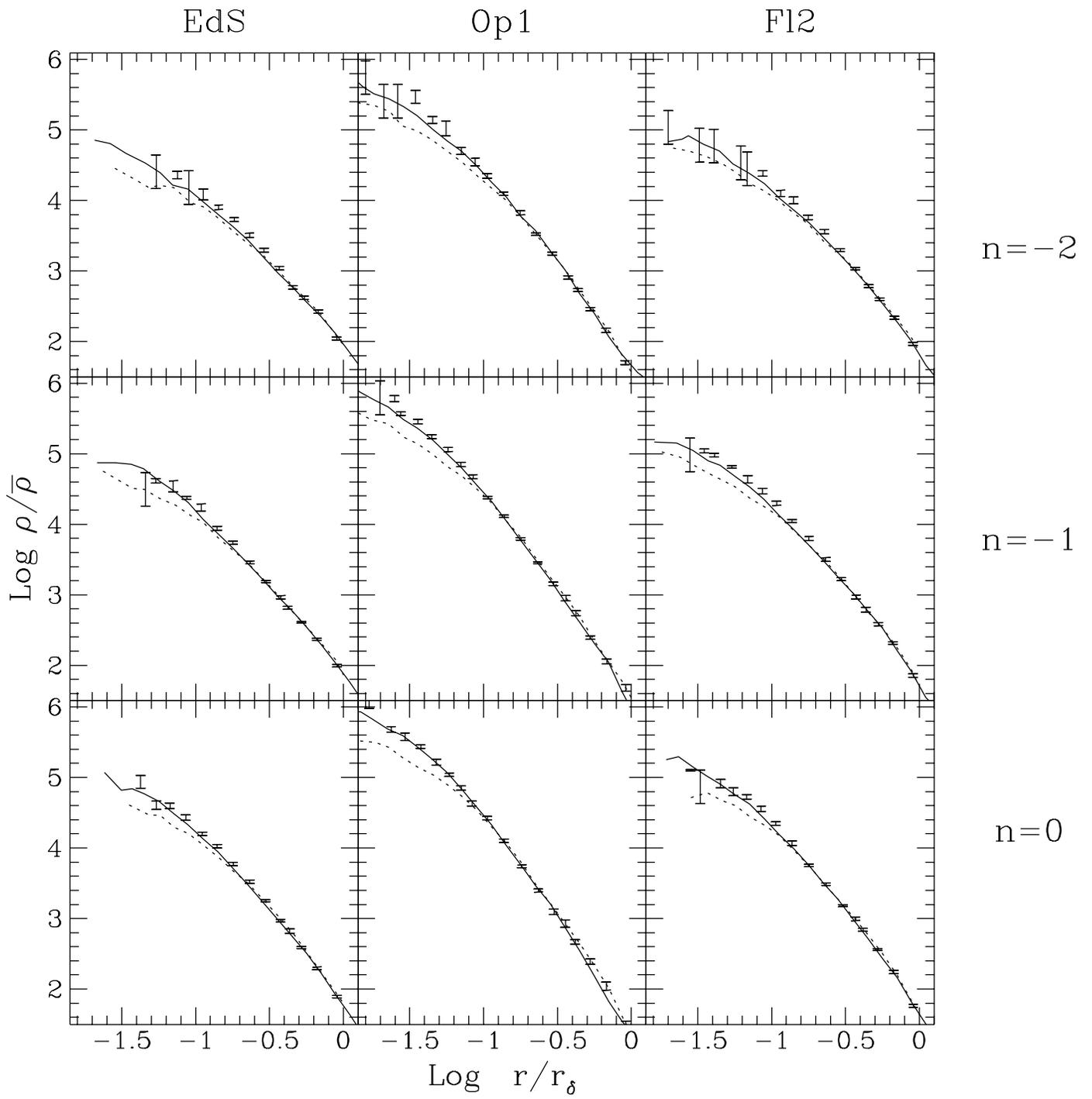

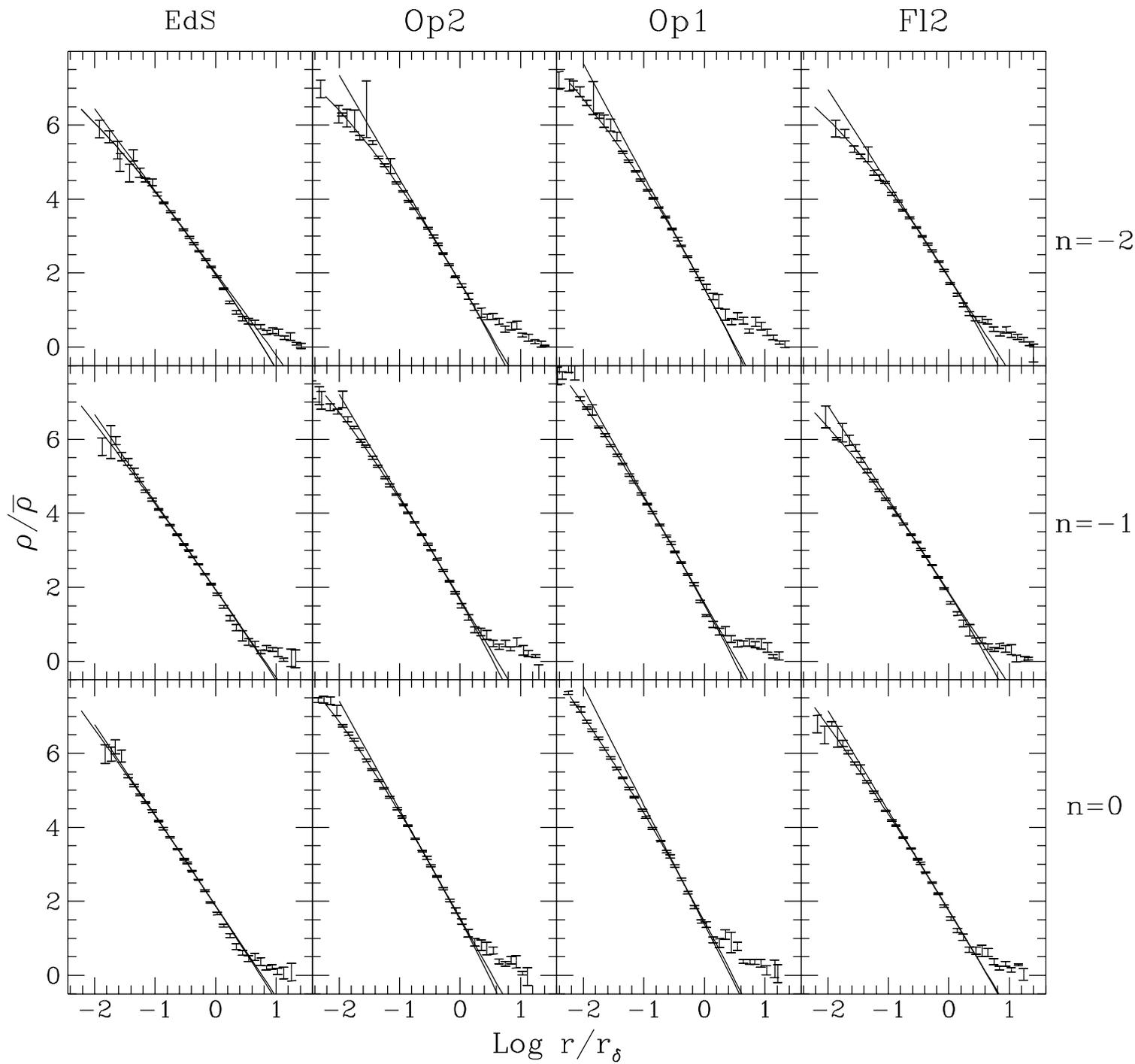

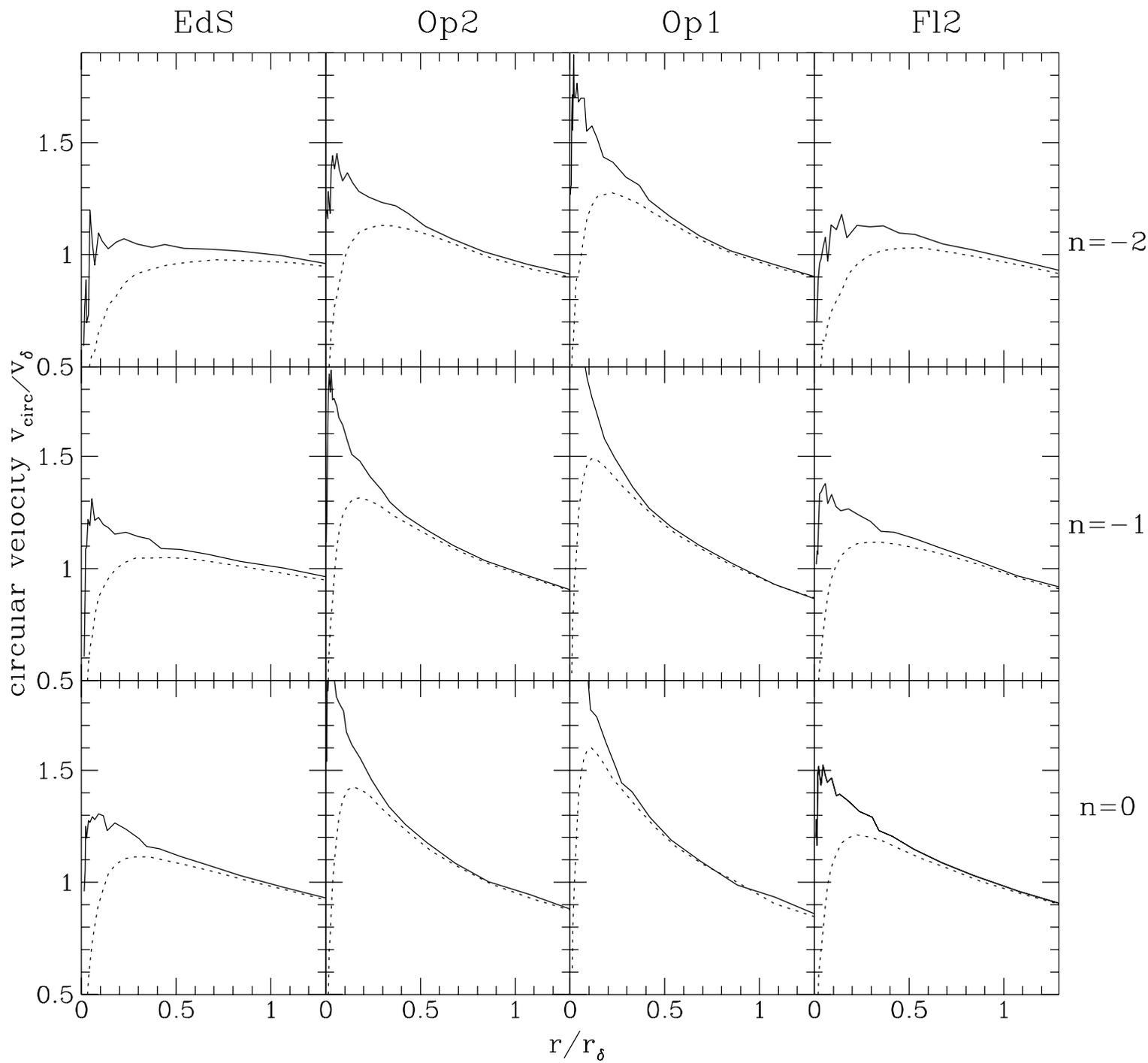

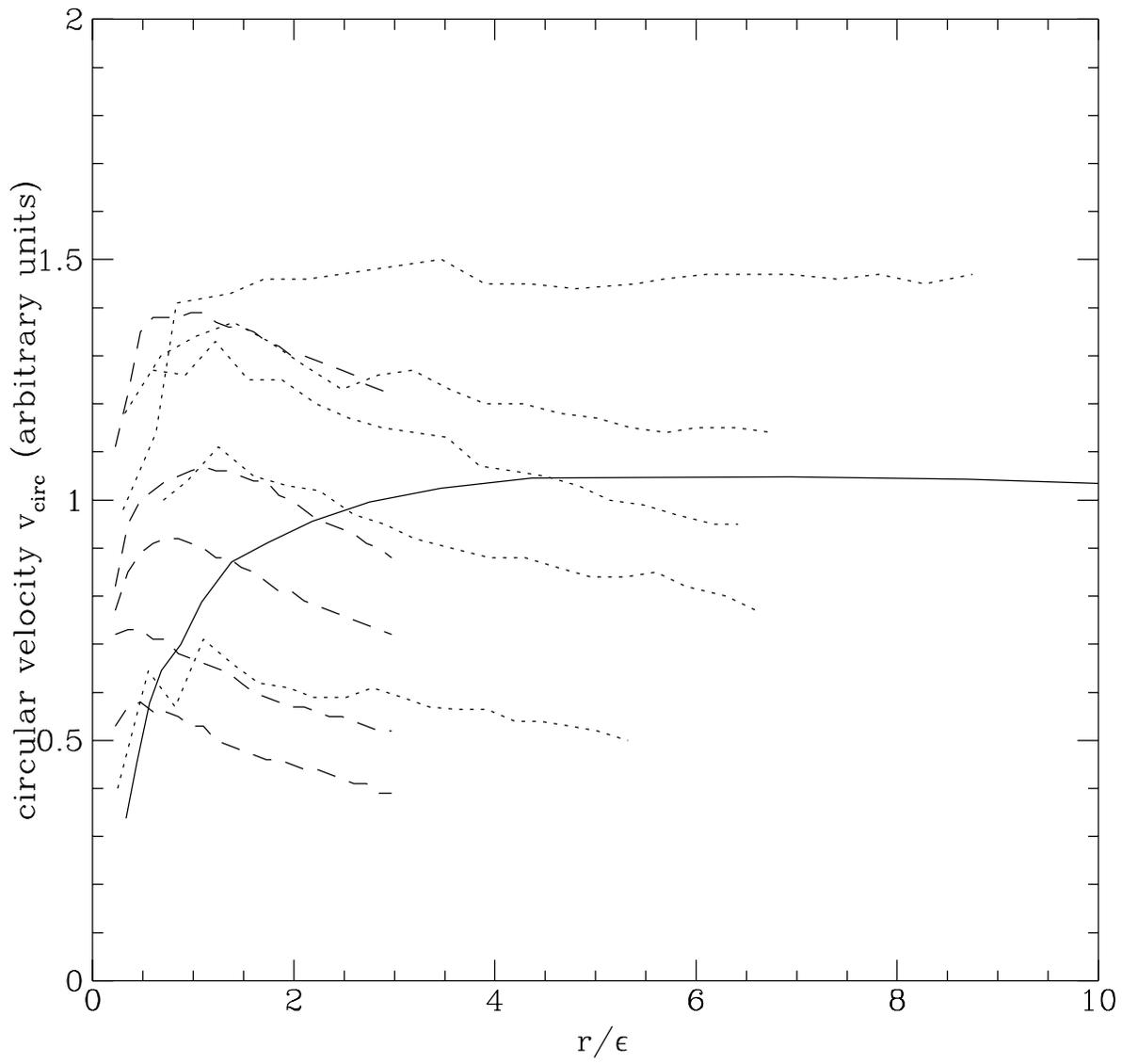

# The Cosmological Dependence of Cluster Density Profiles

Mary M. Crone, August E. Evrard

Physics Department, University of Michigan, Ann Arbor, MI 48109

and

Douglas O. Richstone

Astronomy Department, University of Michigan, Ann Arbor, MI 48109

## ABSTRACT

We use N-body simulations to study the shape of mean cluster density and velocity profiles in the non–linear regime formed via gravitational instability. The dependence of the final structure on both cosmology and initial density field is examined, using a grid of cosmologies and scale-free initial power spectra $P(k) \propto k^n$. Einstein deSitter, open ($\Omega_o = 0.2$ and 0.1) and flat, low density ($\Omega_o = 0.2$, $\lambda_o = 0.8$) models are examined, with initial spectral indices $n = -2, -1$ and 0.

For each model, we stack clusters in an appropriately scaled manner to define an average density profile in the non–linear regime. The profiles are well fit by a power law $\rho(r) \propto r^{-\alpha}$ for radii whereat the local density contrast is between 100 and 3000. This covers 99% of the cluster volume. We find a clear trend toward steeper slopes (larger $\alpha$'s) with both increasing $n$ and decreasing $\Omega_o$. The $\Omega_o$ dependence is partially masked by the $n$ dependence; there is degeneracy in the values of $\alpha$ between the Einstein deSitter and flat, low density cosmologies. However, the profile slopes in the open models are consistently higher than the $\Omega = 1$ values for the range of $n$ examined. Cluster density profiles are thus potentially useful cosmological diagnostics.

We find no evidence for a constant density core in any of the models, although the density profiles do tend to flatten at small radii. Much of the flattening is due to the force softening required by the simulations. An attempt is made to recover the unsoftened profiles assuming angular momentum invariance. The recovered profiles in Einstein deSitter cosmologies are consistent with a pure power law up to the highest density contrasts ($10^6$) accessible with our resolution. The low density models show significant deviations from a power law above density contrasts $\sim 10^5$. We interpret this curvature as reflecting the non scale–invariant nature of the background cosmology in these models. These results are at the limit of our resolution and so should be tested in future using simulations with larger numbers of particles. Such simulations will also provide insight on the broader problem of understanding, in a statistical sense, the full phase space structure of collapsed, cosmological halos.

## 1. Introduction

There is an intriguing possibility that we can constrain cosmological parameters with the properties of galaxy clusters. The hope of deciphering cosmological parameters is rooted in the dynamical youth of clusters. Clusters are so large that, except within their cores, only a few crossing times have taken place in a Hubble time. Unless violent relaxation and phase mixing are extremely efficient, the present structure of clusters should retain information on the initial conditions and evolutionary history that went into forming them. Two specific features which should contain cosmological information are cluster mass density profiles and substructure (Forman & Jones 1982; Quinn, Salmon & Zurek 1986; Fitchett & Webster 1987; West, Dekel & Oemler 1987; Richstone, Loeb & Turner 1992; Evrard *et al.* 1993).

Surprisingly, the question of whether cluster mass density profiles carry an unambiguous cosmological signal remains unresolved, even in the well–studied case of structure formation from Gaussian random density inhomogeneities. Simple gravitational collapse has been shown to produce an $r^{-4}$ profile, while shallower profiles of $r^{-3} - r^{-2}$, in better agreement with observations, can be produced by secondary infall of surrounding material (Gott 1975; Gunn 1977). Self-similar solutions were found by Fillmore & Goldreich (1984) and Bertshinger (1985), who found a profile of $r^{-2.25}$ for the case of accretion onto a point mass perturber in a flat universe. Hoffman & Shaham (1985) added calculations including both open and flat universes, assuming scale–free initial perturbation spectra $P(k) = A_n k^n$ with $-3 < n < 4$. Using a spherical shell approximation to model the collapse of peaks in the initial density field, they recovered Bertshinger's $r^{-2.25}$ profile for $\Omega = 1, n = 0$. They also showed that, in an open universe, logarithmic slopes steepen with $r$ because of the changing $\Omega$ as subsequent mass shells fall in. The limiting $r^{-4}$ profile is reached as $\Omega$ goes to zero. These calculations were refined by Hoffman (1988) who examined the evolution of spherical shells centered on peaks in the initial density field. Early simulations of a core plus infalling shells qualitatively supported these calculations (Gott 1975; Dekel, Kowitt & Shaham 1981; Pryor 1982).

The simplifying assumptions (*e.g.*, spherical symmetry) necessary in the analytic and semi–analytic work cited above casts some doubt on the practical value of such analysis. To overcome this shortcoming, three–dimensional large–scale structure simulations incorporating tens of thousands of particles were pursued in the mid– to late 1980's. Unfortunately, the results were sometimes conflicting. Quinn, Salmon & Zurek (1986, hereafter QSZ) demonstrated a link between the density profiles of collapsed objects and the initial fluctuation spectrum for Einstein deSitter models. However, in a similar study, West, Dekel & Oemler (1987) concluded the opposite — that collapsed density profiles were insensitive to the initial spectrum of fluctuations. Efstathiou *et al.* (1988, hereafter EFWD) examined gravitational clustering in Einstein deSitter cosmologies from scale free initial power spectra with $n = -2, -1, 0$ and 1. Their simulations exhibited density profiles steepening with increasing $n$, which supported the work of QSZ. Finally, Warren *et al.* (1992) saw the same steepening behavior in runs which spanned the same range of $n$ as EFWD, but performed with $\sim 10^6$ particles rather than $32^3$. With regard to the dependence on the density parameter $\Omega_o$, both West *et al.* and Zurek, Quinn & Salmon (1988) demonstrated that mass density profiles steepened in low $\Omega_o$ models, in agreement with the analytic expectations of Hoffman & Shaham (1985, hereafter HS).

None of these studies was designed to systematically measure profiles as a function of both $n$ and $\Omega_o$, or provide estimates of the uncertainty of quoted slopes. To address this, we have performed a set of $64^3$ particle N–body simulations covering a grid of cosmological models and initial, scale–free fluctuation spectra. In this paper, we demonstrate clear links between the underlying cosmology and final, non–linear, cluster density profiles. We will examine the issue of substructure in a later paper.

The choice to study scale–free spectra, rather than specific cosmogonic models such as cold (CDM) or cold+hot dark matter (CHDM), is motivated by their generic nature. The exact shape of the fluctuation spectra of specific models is parameter dependent. In general, one must at least choose a value of the Hubble constant $h$ (where, as usual, $h = H_o/100$ km s$^{-1}$ Mpc$^{-1}$) and the baryon fraction $\Omega_b$, and often there are other parameter choices such as the ratio of mass in hot and cold dark matter for CHDM. For spectra which do not possess sharp characteristic features, a power law will represent a useful approximation over a finite range of scales. For example, CDM on the scale of clusters can be reasonably well approximated by an $n = -1$ spectrum. (On the other hand, a model such as the primeval isocurvature baryon model (Peebles 1987) would not be well approximated in this way on scales larger than clusters, because of the peak in the power spectrum reflecting the Jeans length at decoupling in this model.) Scale–free models thus provide a laboratory for investigating issues applicable to a wide variety of cosmological models.

We find the cluster density profiles are well fit by power laws over a large fraction ($\sim 99\%$) of their outer volume whereas the structure of the central few percent of the mass generally differs from the outer regions; the profiles turn over to shallower slopes at very small radii. This is not surprising for low density models, which have intrinsic scales to break self–similarity in the clustering hierarchy. In an Einstein–deSitter universe, however, the scale–free dynamics of the problem leads to the naive expectation that non–linear density profiles are likely to be pure power laws. Interpreting the turnover at small radii requires careful consideration of numerical resolution. Specifically, force softening employed to suppress two–body relaxation on small scales will generate a turnover which is purely numerical in origin. We attempt to correct for the effects of softening by using an adiabatic invariant approach similar to that used by Blumenthal *et al.* (1986) and Ryden & Gunn (1987) in studying halo response to galactic disks. We examine the "recovered" profiles for signs of curvature at small radii.

In §II, we describe our methods and terminology. In §III, we present power law fits to mean density profiles and analyze the averaged velocity structure of the clusters. We examine the inner profiles using our correction procedure and higher order fits in §IV. We summarize our results and discuss implications in §V.

## 2. Simulations and Analysis

### 2.1. Initial Conditions and Cosmological Models

Ideally, one would like to have a one–to–one correspondence between the assumed structure formation model and the non–linear profiles of collapsed objects. In reality, some level of degeneracy is likely to exist between the initial power spectrum and the assumed cosmological model. In the context of the spherical shell model (Peebles 1980; HS), the final density structure is controlled by the rate of accretion of new mass shells. This rate is determined by both the shape of the initial spectrum and the time dependence of the Hubble parameter $H(t)$ (*e.g.*, Gunn & Gott 1972), which is controlled by the cosmological model. To explore the competing effects of spectral index $n$ and cosmology, we performed simulations in an Einstein–deSitter universe, open Friedmann–Lemaitre models with $\Omega_o = 0.2$ and $\Omega_o = 0.1$, and a flat, low density model with $\Omega_o = 0.2, \lambda_o = 0.8$ where $\lambda_o = \Lambda/3H_o^2$ parameterizes the cosmological constant. For convenience, we will label these cosmologies as EdS, Op2, Op1 and Fl2, respectively, as summarized in Table 1.

We assume a Gaussian random, scale–free initial density field specified by power spectrum $P(k) \propto k^n$. We examine spectral indices $n = -2, -1$ and $0$, a range which encompasses the likely shape of the power spectrum on scales between galaxies and clusters of galaxies (Henry & Arnaud 1990; Fisher *et al.* 1992; Feldman *et al.* 1993, Peacock & Dodds 1993). We generate a random realization of each spectrum by sampling $64^3$ random amplitudes and phases in the Fourier space of a periodic cube of side $L$. Figure 1 shows the realized power spectra, calculated from our initial density fields, are close to scale-free from the fundamental wavenumber $k = 1$ to the Nyquist wavenumber $k = 32$. There is some noise present in the spectra at low $k$ due to the finite size of the simulation volume. The spectra are normalized such that the *rms*, linear evolved amplitude of fluctuations in a top–hat sphere of radius $L/8$ is unity. The Zel'dovich approximation is used to generate particle positions and velocities from the initial, linear amplitude density field (Efstathiou *et al.* 1985) scaled down in amplitude by a factor 16. This sets the starting epoch of the simulations to be $z_i = 15$ (EdS), 45 (Op2), 77 (Op1) and 22 (Fl2).

We use a P$^3$M code (Efstathiou and Eastwood 1981) with a $128^3$ Fourier mesh (for the PM part) and $64^3$ particles to evolve the systems into the non–linear regime. A Plummer law is used to minimize two-body interactions

$$F_{ij} = -\frac{Gm^2 r_{ij}}{(r_{ij}^2 + \epsilon^2)^{3/2}} \qquad (1)$$

where we have set $\epsilon$ to one–eighth of a Fourier mesh cell, or $L/1024$. The additional short–range, particle–particle correction in P$^3$M is of vital importance in extending the dynamic range in length and density. The softening in a PM code is typically a few mesh cells in size (Hockney & Eastwood 1981), so a PM code having force resolution comparable to the runs described here would require at least $2048^3$ cells.

## 2.2. Mean Profile Analysis

At the final epoch, clusters are identified using a friends–of–friends linking algorithm, with linking parameter set to 0.15 times the mean interparticle separation. This should pick out particles in regions where the local density contrast is $\sim 500$. We limit our analysis to the 35 most massive groups found in each run. The clusters so found are typically resolved by a few thousand particles, with a range spanning from 300 to 11000. For each cluster, the position of the most bound particle is used as a center, about which radial profiles of density and velocity are calculated in bins each containing 20 particles. We are interested in obtaining a characteristic, or "average", profile shape for each cosmology. To obtain this, we use dimensionless variables tied to the overdensity defining the cluster population.

Specifically, for each cluster, in place of radius $r$, we use the radial variable $r/r_\delta$, where $r_\delta$ is defined as the radius of the sphere within which the mean interior density constrast of the cluster is equal to $\delta$. We chose $\delta = 300$, a value believed to demarcate the interior, near hydrostatic from the outer, infalling regimes. We show below that the velocity profiles indicate this choice is reasonably correct. For each simulation, mean cluster profiles in density and velocity are defined by averaging data in bins of width 0.1 in (decimal) $\log(r/r_\delta)$. The variance in each bin is used to compute the error in the mean value in the usual way.

The interpretation of an "average" profile is straightforward if clusters within a given model are strictly self–similar. The assumption of self–similarity is that density profiles of collapsed objects $\rho(r,t)$ examined at some time $t$ can be re–written

$$\rho(r,t) = \rho(\lambda r_c(t), t) \equiv D(\lambda)\bar{\rho}(t) \qquad (2)$$

in terms of a product of a dimensionless function $D(\lambda)$ and the mean background density $\bar{\rho}(t)$ (*e.g.*, Bertschinger 1985). The dimensionless radius $\lambda \equiv r/r_c(t)$ expresses the radius of a cluster in terms of some characteristic radius, such as the turnaround radius of the cluster at time $t$. Our use of $r_\delta(t)$ as a characteristic radius is simply a convention, allowing us to use units in which the radius of a cluster is unity. One could employ the turnaround radius $r_{ta}(t)$ via a change of variable $\lambda \longrightarrow \lambda' \equiv (r_\delta(t)/r_{ta}(t))\lambda$ since $r_\delta(t)/r_{ta}(t)$ is a (model dependent) constant for a self–similar profile.

There is currently no proof that clusters grown in any cosmological context must be self–similar. That is, the use of eqn. (2) with a single function $D$ for all clusters in a given cosmogony is, at this point, an *assumption*. With scale–free initial conditions, it is reasonable to suspect that the only possible scales imparted during gravitational clustering must be cosmological in origin and, therefore, dependent only on epoch. In that case, clusters examined at a fixed epoch should be self–similar, in the sense that big clusters should simply look like magnified versions of little clusters. On the other hand, the dynamical histories of big and little clusters may be different, perhaps with the former linked to high peaks in the linear density field and the latter generated more by shear flows (Bertschinger & Jain 1993). Different dynamical histories could lead to different non–linear structural properties. However, both because we are interested in rich clusters of galaxies and because of our limited dynamic range, we are concentrating on the high mass end of the cluster mass function. We have investigated whether the most massive clusters in each simulation obey self–similarity by computing average density profiles for the clusters ranked $1-10$, $11-20$ and $21-30$ by mass. The results for $\Omega = 1$, $n = -1$ are shown in Figure 2, results from the other models are very similar. There is good evidence for self–similarity among the most massive objects. Very high resolution studies will be required to determine whether self–similarity holds over a wider range of masses.

In an matter dominated, Einstein deSitter universe, the scale factor and Hubble parameter are power–laws in time, reflecting the fact that there are no characteristic epochs in such a cosmology. Since neither gravity nor the initial conditions have a preferred scale, one might suspect that the non–linear density profiles of collapsed objects should be power law in form. The reasoning goes something like this. If the profiles in the non-linear regime of the EdS models were NOT power law, then there must be at least one characteristic scale, say $\lambda_*$, which describes the position of a bend or some other feature in the density profile; *i.e.*, the self–similar density profile of equation (2) above must be written

$$D(\lambda) \;=\; f(\lambda/\lambda_*). \tag{3}$$

One can argue that $r_\delta$ is defined by invoking the virial theorem to separate the infall from the hydrostatic regimes. The question that is raised immediately is what physics defines $\lambda_*$? What scales are there in the problem that one can tie $\lambda_*$ to? We feel there is no obvious answer to that question, which leads to the "naive" point of view that there is no such scale and that therefore $\lambda_*$ doesn't exist. The only possible form for $D(\lambda)$ is then a power law.

In practice, one must be concerned that there are numerical scales introduced into the problem; namely, the gravitational softening $\epsilon$ and the Nyquist frequency and fundamental mode of the initial power spectrum. Although we attempt to correct for the former below, we stress the need for future simulations with extended dynamic range to address these issues directly.

## 3. Dependence on Spectral Index and Cosmology

### 3.1. Density Structure

The average density profiles for the 35 most massive clusters in each simulation are shown in Figure 3a. Generic features of the profiles include a central region of progressively steepening slope, leading to an intermediate, nearly power law regime which extends to $\rho/\bar{\rho} \sim 100$. The outer parts of the cluster first steepen in an infall regime extending to $\rho/\bar{\rho} \sim 3-10$, then turn up to meet the mean background value. We have fit the profiles to a power law

$$\frac{\rho}{\bar{\rho}} \;=\; \Delta \left(\frac{r}{r_\delta}\right)^{-\alpha} \tag{4}$$

in the region where the local density contrast is in the range $100 < \rho/\bar{\rho} < 3000$. The fits are shown as the solid lines in Figure 3a and values of the slope $\alpha$ are listed in Table 2. In generating the mean profiles, we have excluded particles within $4\epsilon$ of each cluster center. As discussed in the next section, the effect of force softening is less than 13% for the particles included in this analysis.

The range in density within which we fit is bounded from below by the edge of the hydrostatic portion of the cluster, *i.e.*, the regime where the mean radial velocity is zero, and is bounded from above by our numerical resolution. From an observational point of view, this regime is not terribly restrictive, since it corresponds to the outer $\sim 80\%$ of the cluster radius, or 96% of the surface area and better than 99% of the cluster volume. For example, in a cluster as massive as Coma (White *et al.* 1993), the density range for which we are fitting would cover roughly the range $0.3 - 1.5 \, h^{-1}$ Mpc in radius. Observationally, the binding mass structure in this regime can be probed by X–ray observations and also by weak gravitational lensing (Kaiser & Squires 1993). We attempt to extend the range of fits to higher density contrast in the next section.

Values of $\chi^2$ for the fits (Table 2) verify that, within this regime, most of the profiles are well fit by power laws. However, residuals from the fits, shown in Figure 3b, show evidence for curvature in the density profiles. The maximum deviation in this range is $\sim 40\%$. The causes of this curvature are discussed further below.

Values of the fitted slopes $\alpha$ are given in Figure 4 as a function of $n$ for the different cosmological models. Trends with both $n$ and $\Omega_o$ plainly exist: slopes steepen with increasing $n$ and decreasing $\Omega_o$. Values for the flat, $\Omega_o = 0.2$ model fall between those of $\Omega_o = 0.2$ and $\Omega_o = 1.0$. As expected, it is not possible to unambiguously determine $\Omega_o$ from cluster density profiles because of the degeneracy with $n$. However, if a cosmological constant were ruled out, one could hope to differentiate between $\Omega_o \lesssim 0.2$ and $\Omega = 1.0$, since our fits from 35 clusters indicate the mean slopes differ by at least $4\sigma$ over the range of $n$ tested here.

It is apparent from Figure 4 that none of the profiles agree with an isothermal profile slope of $\alpha = 2$. At first glance, this result contradicts the qualitative conclusion of QSZ and EFWD, who claimed to find flat rotation curves

$$v_{rot}(r) \;=\; \sqrt{\frac{GM(<r)}{r}} \;\propto\; r^{1-\alpha/2} \tag{5}$$

for $n = -1$ and $-2$ in an EdS cosmology. We argue below that force softening on small scales in their experiments was likely to be responsible for the shallower density profiles they observed. Also, the magnitude of the velocity gradient is small; for our $n = -1$ result, $\alpha = 2.33 \pm 0.04$, only an 11% drop in $v_{rot}$ occurs over a factor 2 in radius.

## 3.2. Velocity Structure

Although more difficult to probe observationally, the velocity structure of the clusters is also of interest, since it provides crucial information on the question of hydrostatic equilibrium needed to interpret the density structure discussed above. We express velocities in terms of $v_\delta$, the circular velocity at $r_\delta$

$$v_\delta = \sqrt{\frac{GM(<r_\delta)}{r_\delta}} = \sqrt{\frac{\Omega_o \delta}{2}} H_o r_\delta \qquad (6)$$

where the last expression is based on the defining criterion of $r_\delta$. Figure 5 shows the mean radial velocity profiles of clusters in each of the simulations. The dotted line in each panel shows the Hubble flow.

The data in Figure 5 support the use of our density contrast criterion to demarcate the inner, hydrostatically supported regime from the outer, infall regime. The mean radial velocities within $r_\delta$ are near zero, indicating hydrostatic equilibrium in this region. As expected from linear growth analysis (Peebles 1980), the infall regime is strongest in the EdS models and weakest in the OP1 runs. The relative lack of infall in low $\Omega_o$ universes reflects the fact that gravitational infall slows and eventually shuts off as $\Omega \to 0$. Within a given cosmology, the infall signature is largest for $n = -2$. This arises because flatter initial spectra produce larger present mass accretion rates, equivalent to more recent cluster formation (Lacey & Cole 1993). It is interesting that *all* the cosmologies show evidence for infall when $n = -2$.

The transition from infall to stably clustered regimes is not perfectly sharp. For most models, the velocity is slightly negative just inside $r_\delta$. The settling occurring at these radii is likely to be responsible for the steepening of the density profiles near $r_\delta$ seen in Figure 3b. As can be seen from the continuity equation evaluated just outside the hydrostatic regime

$$\frac{\partial \rho}{\partial t} = -\frac{1}{r^2}\frac{\partial(r^2 \rho \bar{v}_r)}{\partial r} \propto -\rho \frac{\partial(\bar{v}_r)}{\partial r} \qquad (7)$$

the density is not constant if there is a net infall, but is still increasing.

The *rms* three–dimensional velocity dispersion $\sigma$, normalized to $v_\delta$, is shown in Figure 6. In calculating $\sigma$, the radial dispersion at a given radius was calculated about the mean value at that radius, rather than about zero. For fixed $n$, the velocity dispersion drops more quickly with radius in low density models. Within a given cosmology, the dispersions drop more quickly as $n$ is increased. As with the density profiles, there is degeneracy in the velocity dispersion profiles such that low density models with flat initial spectra appear similar to EdS models with steep spectra. The orbital anisotropy parameter

$$A(r) = 1 - \frac{\sigma_t^2}{\sigma_r^2}$$

is shown in Figure 7, where $\sigma_r$ is the radial velocity dispersion measured about the mean $v_r$ and $\sigma_t$ is the one–dimensional tangential dispersion. All models exhibit a tendency toward radial orbits, with the anisotropy increasing weakly with radius to values $A(r) = 0.2 - 0.5$ at $r = r_\delta$. At a given $r/r_\delta$, the anisotropy is somewhat larger in the open models and for steeper initial spectra.

## 4. The Inner Density Profile and Expected Rotation Curves

The fits performed in the previous section covered a large fraction of the cluster volume, but excluded what is arguably the most interesting region — the very center or "core" of the clusters. There has been considerable recent interest in the structure of cluster cores due to constraints derived from arcs produced by gravitational lensing (Tyson, Valdes & Wenk 1990; Miralde–Escudé 1993; Wu & Hammer 1993). Fits to a binding mass profile of the truncated isothermal form

$$\rho(r) = \frac{\rho_o}{1 + (r/r_c)^2} \qquad (8)$$

for clusters observed to have strongly lensed arcs indicate that the core radius $r_c$ for the binding mass distribution is surprisingly small. Wu & Hammer (1993) require $r_c$ for the dark matter to be a factor 10 below that of the X–ray gas, implying $r_c \lesssim 50\,h^{-1}$ kpc. Such small values come as no shock to N-body simulators, since fits to eqn. (8) with *resolved* values of $r_c$ have never been seen in simulations of hierarchical clustering from popular "bottom–up" models like CDM (Dubinski & Carlberg 1991 and references therein). The only simulations which produced resolved, constant density cores were those of McGlynn (1984), whose "warm start" initial conditions imposed a minimum phase space density on the structure of the clusters. The fault lies in the assumption that equation (8) (or similar variants with constant central density) should describe the density structure of the binding mass profiles.

Unfortunately, there is no firm analytic framework which connects the non–linear, "stellar dynamical" description of collapsed regions to the linear, initial density fluctuations. There are, of course, theoretical prejudices concerning the shape of the central density profile. We argued above reasons to expect a power law density profile to extend to arbitrarily small scales for the case of clustering from scale–free spectra in an Einstein–deSitter cosmology. Essentially, a flat, matter dominated universe with initial power spectrum $P(k) \propto k^n$ imposes no scales on gravitational collapse, so we might expect power law profiles interior to some caustic surface separating the hydrostatic and infall regimes. Hoffman (1989) offers an alternative point of view based on the expected shape of the linear density profiles around peaks. In this picture, the final structure of objects is closely tied to the existence and height of peaks on the range of mass scales of interest. Power law profiles are not generally expected in collapsed objects.

In an open universe, there is an obvious scale, the radius of curvature of the universe, which dictates when the universe begins to deviate significantly from a flat geometry. HS argued that collapsed density profiles steepen as $\Omega_o$ decreases, so one might expect the final structure of low $\Omega_o$ clusters to exhibit progressively steeper slopes at large radii. In such a picture, one might expect that the cores of such clusters may have density profiles closely those obtained in an $\Omega = 1$ universe.

### 4.1. Correcting for Softening — A Recovery Procedure

To examine the structure of the profiles at high density contrast in our simulations requires that we somehow correct for the effects of force softening. The analysis in the previous section simply ignored the inner regions of the clusters, where softening effects are significant. As shown in Figure 8 (solid line), the ratio of the softened to unsoftened forces $F_s/F$ for a point mass is greater than 0.9 at distances $r > 4\epsilon$. For spherically symmetric, extended objects with power–law density profiles, the ratio of $F_s/F$ found by numerical integration is also shown in Figure 8. For an object with $\alpha = 2.0$, a lower bound to the slopes in Figure 4, $F_s/F \geq 87\%$ at $r > 4\epsilon$. Because simulated clusters do not reach infinite density at $r = 0$, a more realistic density profile is cut off to a constant density at $r \lesssim \epsilon/10$. Including such a cutoff leads to a negligible change in $F_s/F$ at $r = 4\epsilon$ for $\alpha < 3$. However, imposing the cutoff allows calculation of $F_s/F$ for $\alpha \geq 3.0$, which would be undefined due to the diverging mass integral without a cutoff. This case is also shown in Figure 8. As expected, the steeper density profiles tend to look more "pointlike", so the range of interest in $F_s/F$ is spanned by the $\alpha = 2$ case and the point mass case. At $r = 4\epsilon$, this translates into softened to unsoftened force ratios of $0.87 < F_s/F < 0.92$.

To illustrate the effects of softening on density profiles, Figure 9 displays the density profiles from our standard set of runs, along with the profiles from a set of runs using the same initial conditions but with $\epsilon$ set a factor 2 larger. (For arcane reasons, we did not run the Op2 models with large $\epsilon$.) The runs with larger $\epsilon$ exhibit a stronger turnover at small radii and reach lower central densities than their counterparts with smaller $\epsilon$, consistent with the softer central forces found in the former.

We attempt to correct for the effects of force softening using a procedure similar in spirit to that used by Blumenthal *et al.* (1986) and Ryden & Gunn (1987) in modeling the effect of galactic disk growth on the density structure of dark matter halos. The correction is based on angular momentum invariance in a central force. The force acting on a test particle within a spherical overdensity is central, so the particle's angular momentum about that center is independent of the magnitude of the force. In particular, if the interparticle force is softened, the central force is weaker, but the distribution of angular momenta is conserved. Due to the softening, each particle is, on average, at a larger radius. If we were to adiabatically "harden" the interparticle forces (whilst simultaneously increasing the number of particles in the system to avoid the ill effects of two–body relaxation), the particle orbits apocentric and pericentric distances would shrink in such a way as to conserve its angular momentum. The effects are clearly largest at small radii, where the effects of softening are largest. The result should be a density profile steepened in the inner regions with respect to the original, softened version.

The actual procedure we employ is the following. To each bin $i$ in our spherically averaged densities, we define a characteristic specific angular momentum based on circular orbits

$$j_{\theta,i} = v_{rot,i} r_i = \sqrt{r_i^3 F_\epsilon(r_i)/M_i} \qquad (9)$$

where $M_i$ is the mass on the $i^{\text{th}}$ shell and $F_\epsilon(r_i)$ is the softened force at radius $r_i$. The "correct" radius $r_{c,i}$ is that which conserves the specific angular momentum when $\epsilon = 0$

$$\sqrt{r_{c,i}^3 F_{\epsilon=0}(r_{c,i})/M_i} \;=\; \sqrt{GM(<r_i)r_{c,i}} \;=\; j_{\theta,i} \qquad (10)$$

where $M(<r_i)$ is the mass interior to the $i^{\text{th}}$ shell. The calculation is particularly simple given that the unsoftened force depends only on the interior mass, which does not change (i.e., there is no shell crossing). When a shell is moved from $r_i$ to $r_{c,i} < r_i$, the volume it occupies shrinks, and the density correspondingly increases.

To test the accuracy of this procedure empirically, we applied it to the runs with large softening, shown in Figure 9, to see if it could reproduce the results obtained with our original softening. That is, rather than $\epsilon \to 0$, we are attempting $\epsilon \to \epsilon/2$. The data points with error bars in Figure 9 show the results of this procedure. The data points, which originally defined the dotted lines of the large $\epsilon$ solutions, have been raised to higher densities at small radii. The agreement between the "recovered" large $\epsilon$ profiles and the small $\epsilon$ profiles is quite good. In the Op1 runs, the correction can be as much as a factor of 3 in density at small radii. At large radii, the profiles remain unchanged, as one would expect given the small effect that softening has for either set of runs.

This empirical test gives us confidence to apply our procedure to recover unsoftened ($\epsilon = 0$) profiles from the original data in Figure 3a. The correction procedure was applied on an individual cluster basis, and resultant average profiles calculated in a manner identical to that used in the previous section. The results are shown in Figure 10. Note the expanded scale with respect to that used in Figure 3a.

## 4.2. Beyond Power Law Fits

The density profiles above a local density contrast $\rho/\bar{\rho} = 100$ in Figure 10 show that the EdS models appear close to pure power laws, while the low density models do not. To measure the departure from power law behavior, we fit the profiles to a quadratic form in $\log(\rho/\bar{\rho})$ against $\log(r/r_\delta)$ which implies a density law of the form

$$\frac{\rho}{\bar{\rho}} \;=\; \Delta' \left(\frac{r}{r_\delta}\right)^{-(\alpha + \beta \log(r/r_\delta))} \qquad (11)$$

where decimal logarithms are used throughout. Here $\alpha$ is the logarithmic slope at $r = r_\delta$, and $\beta$ is the change in $\alpha$ over each decade in $r$. Table 3 give the values of $\alpha$ and $\beta$ obtained from fits of the data in Figure 10 above a local density contrast of 100. The low density models all have non-zero values of $\beta$, with a typical value of $\beta \sim 0.2 \pm 0.03$. There is not enough information to distinguish trends in values of $\beta$ with cosmology or spectral index. The EdS runs with $n = -1$ and 0 are consistent with $\beta = 0$ while the $n = -2$ run has $\beta = 0.17 \pm 0.05$. We suspect that the strong infall occurring in the regime $100 < \bar{\rho} < 500$ (see Figures 3a and 5) is forcing the value of $\alpha$ up in this fit, since it is sensitive to the density structure around $r = r_\delta$. (Compare the values of $\alpha$ in Tables 2 and 3.) A non–zero value of $\beta$ results because of the shallower profile which exists in the hydrostatic regime at higher density contrasts.

The three parameter description above is not intended to be true physical description over an arbitrarily large range in radius. This can be seen by the fact that, for models with positive $\beta$, the density profile at very small radii will eventually turn over and the local logarathmic slope become arbitrarily large and positive. A more physically plausible form might be a four parameter, two–power law form such as

$$\frac{\rho}{\bar{\rho}} = \Delta'' \left(\frac{r}{\lambda_c r_\delta}\right)^{-\eta} \left(1 + \left(\frac{r}{\lambda_c r_\delta}\right)^2\right)^{-\gamma/2} \qquad (12)$$

where the profile changes from having slope $-\eta$ to $-(\eta + \gamma)$ around a critical radius $r = \lambda_c r_\delta$. The motivation for this form comes from considering that the inner regions of the clusters in low density models collapse early, when the background cosmology is much closer to Einstein–deSitter. If late–infalling shells do not strongly affect the central structure, then the inner density profiles of these models should resemble those formed in EdS cosmologies. In that case, the values of $\eta$ found from the above form should agree with the values of $\alpha$ from the pure power fits of the EdS models in Table 2. Table 4 shows the values of the parameters obtained from fitting eqn. (12) to the density profiles of Figure 10. The values of $\eta$ in the low density models consistently fall $1 - 2\sigma$ below the EdS values of $\alpha$ from Table 2. It is not clear why this difference arises and whether it is significant, since the fits are rather poor. It could be an artifact of our density recovery procedure, or it could be that the EdS profiles are slightly steeper due to infall.

We conclude from this analysis that, in the hydrostatic regime, the density profiles of clusters in the EdS models are consistent with pure power laws, the slope of which depends on the spectral index as shown in Table 2 and Figure 4. In addition, the lack of scale invariance in low density cosmological models imprints curvature in the non–linear density profiles of clusters. The inner slopes of the low density clusters are marginally consistent with the EdS slopes. These conclusions should be tested with higher resolution N–body experiments in the near future.

### 4.3. Rotation Curve Shapes

To compare with the previous work of QSZ and EFWD, we replot in Figure 11 the density profiles data of Figures 3a and 10, expressed in terms of an equilibrium rotation speed defined in eqn. (5) above, normalized to the characteristic velocity $v_\delta$ given in eqn. (6). Both QSZ and EFWD claimed evidence for flat rotation curves from scale–free spectra with $n = -2$ and $-1$ in EdS models. Both of these studies were done using smaller numbers of particles than we are using, and the analysis of rotation curves was performed very near the spatial resolution limit of the simulations. Neither study attempted to systematically correct for softening effects. Warren *et al.*

(1992) display rotation curves for a $n = -1$ EdS experiments in which the softening is changed by a factor of 5. The structure within $4\epsilon$ is noticeably altered, consistent with the results we show here.

The dashed lines in Figure 11 show the rotation curves generated from our raw data. It is clear that, for EdS models with $n = -2$ and $-1$, there is a broad range in radius over which the velocity profile appears flat. When corrected for softening (solid lines), none of the profiles appears perfectly flat, although the $n = -2$ EDS model has a velocity which varies by less than 10% over the entire range plotted.

In Figure 12, we reproduce the circular velocity data from QSZ (their Figure 2) and EFWD (their Figure 15) for the $n = -1$ EdS model. The circular velocity is in arbitrary units while the radius is given in units of the gravitational softening, quoted as 10 kpc for QSZ and $0.05L/64$ for EFWD. (The force law of EFDW was not a Plummer law; we determined an equivalent $\epsilon$ for their models by determining the radius at which their force law deviates from Newtonian by 10% and equating it to $4\epsilon$.) The QSZ data are for individual halos, while the EFWD data are averages in mass bins. Again, the rotation curves appear flat at small radii, then decline at larger radii where the effect of force softening is small. The decline at large radii is consistent with eqn. (5) with our fitted slope of $\alpha = 2.33$, as indicated by the bold line in Figure 12.

To scale the results in Figure 11 to physical dimensions, recall that the mean overdensity within radius $r$ can be expressed via the rotation velocity and cosmological parameters by

$$\delta = \frac{2}{\Omega_o} \left( \frac{v_{rot}}{H_o r} \right)^2 \tag{13}$$

which can be used to solve for the radius in terms of $v_{rot}$ and $\delta$

$$r = \sqrt{\frac{2}{\Omega_o \delta}} \left( \frac{v_{rot}}{H_o} \right) = 163 \, \Omega_o^{-1/2} \left( \frac{\delta}{300} \right)^{-1/2} \left( \frac{v_{rot}}{200 \text{ km s}^{-1}} \right) h^{-1} \text{ kpc.} \tag{14}$$

For a galaxy with $v_{rot} = 200$ km s$^{-1}$, our power law fits performed in §3.1 apply to a range in radius between roughly 30 and $160 \, h^{-1}$ kpc. This region is generally just beyond the region where high quality rotation curves are available.

Our results mildly conflict with the analytic scaling arguments presented by HS. The simulations support the overall trend, predicted by scaling arguments, of steeper density profiles with either increasing $n$ or decreasing $\Omega_o$. However, the values of the density profile slopes are somewhat steeper in the simulations than predicted by HS. In this regard, the simulations agree better with the refined calculations of Hoffman (1988). At any rate, the difference between the analytically predicted and numerically determined slopes is small ($\lesssim 20\%$), making it difficult to pinpoint the underlying cause for the discrepancy. Disagreement between the analytic and numerical work is perhaps not too surprising given both the simplifying assumptions that must be made analytically — namely, spherically symmetric mean peak profiles and the assumption that all mass shells have similar collapse histories — and the introduction of numerical scales within the simulation description. Studies of the dynamical histories of clusters in a subset of these simulations show deviations from the spherical model trajectory (Schreiber & Evrard 1994), although at a fairly modest level. Bertshinger & Jain (1993) have recently emphasized the role of shear in aiding perturbation growth, an aspect lacking in the spherical model. On the other hand, Bernardeau (1993) has calculated exact behavior of the collapse of rare peaks and claims agreement with spherical model evolution.

## 5. Summary and Discussion

We conclude that the density profiles of objects collapsed from scale–free initial power spectra contain information on both the spectral index $n$ and the underlying cosmology. This implies that profiles of the binding mass distribution in a sample of a few tens of clusters could be used to place constraints on the cosmological parameters $\Omega_o$ and $\lambda_o$ and the shape of the fluctuation spectrum near the mass scale of clusters. The degeneracy between $n$ and cosmological parameters (Figure 4) could be broken by using other, independent methods, such as empirical determination of the power spectrum shape using homogeneous, large–scale galaxy samples (Fisher *et al.* 1992; Feldman, Kaiser & Peacock 1994; Peacock & Dodds 1994).

The trend toward steeper density profiles with lower $\Omega_o$ confirms the earlier work of West, Dekel & Oemler (1987) and agrees with the recent gas dynamic simulations of X–ray clusters by Evrard *et al.* (1993). For Einstein deSitter models, our profile slopes are somewhat steeper than the analytic predictions of Hoffman & Shaham (1985) but agree better with the analytic work of Hoffman (1988) and the numerical simulations of Quinn *et al.* (1986) and Efstathiou *et al.* (1988) after softening effects are taken into account. The density profiles at radii near the gravitational softening will be artificially shallow due to the weaker central gravity. We find $\rho \propto r^{-2.20\pm0.04}$ and $r^{-2.33\pm0.04}$ for EdS models with $n=-2$ and $-1$, respectively, whereas the previous work claimed $\rho \propto r^{-2}$. We therefore find no *exactly flat* rotation curves from our models, though extrapolation of Figure 4 would imply that and EdS model with $n\sim -2.5$ would produce $\rho \propto r^{-2}$. It must be remembered that galaxy rotation curves are influenced by the baryonic component of galaxies, which we are ignoring here, and are typically well measured at radii which encompass mean interior density contrasts $\gtrsim 10^5$. Katz & Gunn (1992) modeled individual galaxies with a two–component gas dynamic scheme and found rotation curves which were close to flat, but slightly declining in the outer parts.

We apply a correction procedure assuming angular momentum conservation in an attempt to remove the effects of gravitational softening on the central structure of clusters. Our results indicate that the scale–free nature of clustering in $\Omega=1$ cosmologies with power–law initial spectra imparts no characteristic features in the non–linear density profiles of collapsed objects. That is, the density profiles in the hydrostatic regime above a density contrast $\delta \sim 300$ are consistent with power laws. This is not the case for low density models, which lack scale–invariance in the behavior of cosmological factors with time. However, curvature in density profiles is not an unambiguous signal of low $\Omega_o$, since non–power law density profiles can arise in Einstein deSitter models if the initial power spectrum is not scale–free. Such is the case for the cold dark matter model (Dubinski & Carlberg 1991). Indeed, the difference in the physics controlling perturbation growth during the radiation and matter dominated eras nearly guarantees departure from scale–invariant spectra in realistic models.

We stress the need for higher resolution N–body simulations to address more thoroughly the issue of the shape of the inner density profile. An increase in particle number by a factor of 100 is feasible on parallel machines. For a power law density profile, $\rho \propto r^{-\alpha} \propto M^{-\alpha/(3-\alpha)}$, this improved mass resolution should increase the dynamic range in density by more than four orders of magnitude for values of $\alpha$ larger than 2.

We thank D. Weinberg for participation in the early stages of this project and the referee, Y. Hoffman, for constructive, critical comments. This work was supported by NASA Theory Grant NAGW-2367 and a Faculty Grant from the Horace H. Rackham School of Graduate Studies at the University of Michigan.

**Table 1**

Labels for Cosmological Models

| Label | Model |
|---|---|
| EdS | $\Omega = 1$ |
| Op2 | $\Omega_o = 0.2$ $\lambda_o = 0$ |
| Op1 | $\Omega_o = 0.1$ $\lambda_o = 0$ |
| Fl2 | $\Omega_o = 0.2$, $\lambda_o = 0.8$ |

Table 2. Power Law Fits to Density Profiles

| MODEL | | $\alpha$ | $\chi^2/\nu\ [\nu]$ |
|---|---|---|---|
| $\Omega_o = 1.0$ | | | |
| | $n = -2$ | 2.20± 0.04 | 1.1[5] |
| | $n = -1$ | 2.33± 0.04 | 2.2[5] |
| | $n = 0$ | 2.51± 0.05 | 1.2[4] |
| $\Omega_o = 0.2$ | | | |
| | $n = -2$ | 2.78± 0.06 | 1.4[3] |
| | $n = -1$ | 2.79± 0.06 | 0.9[3] |
| | $n = 0$ | 3.07± 0.08 | 0.2[2] |
| $\Omega_o = 0.1$ | | | |
| | $n = -2$ | 3.09± 0.07 | 0.9[3] |
| | $n = -1$ | 2.99± 0.08 | 0.2[3] |
| | $n = 0$ | 3.25± 0.09 | 1.3[3] |
| $\Omega_o = 0.2, \Omega_o = 0.8$ | | | |
| | $n = -2$ | 2.47± 0.04 | 2.4[5] |
| | $n = -1$ | 2.51± 0.04 | 4.1[4] |
| | $n = 0$ | 2.78± 0.05 | 0.1[3] |

Table 3. Quadratic Fits to Corrected Density Profiles

| MODEL | $\alpha$ | $\beta$ | $\chi^2/\nu$ [$\nu$] |
|---|---|---|---|
| $\Omega_o = 1.0$ | | | |
| $n = -2$ | $2.39 \pm 0.07$ | $0.17 \pm 0.05$ | 0.9[15] |
| $n = -1$ | $2.39 \pm 0.06$ | $0.01 \pm 0.04$ | 1.3[16] |
| $n = 0$ | $2.49 \pm 0.07$ | $0.01 \pm 0.05$ | 0.7[15] |
| $\Omega_o = 0.2$ | | | |
| $n = -2$ | $2.83 \pm 0.05$ | $0.26 \pm 0.03$ | 1.5[19] |
| $n = -1$ | $2.93 \pm 0.04$ | $0.20 \pm 0.02$ | 1.9[21] |
| $n = 0$ | $3.11 \pm 0.05$ | $0.22 \pm 0.02$ | 1.2[20] |
| $\Omega_o = 0.1$ | | | |
| $n = -2$ | $3.08 \pm 0.05$ | $0.27 \pm 0.03$ | 1.0[21] |
| $n = -1$ | $3.04 \pm 0.04$ | $0.17 \pm 0.02$ | 0.8[26] |
| $n = 0$ | $3.09 \pm 0.05$ | $0.16 \pm 0.02$ | 0.9[23] |
| $\Omega_o = 0.2, \Omega_o = 0.8$ | | | |
| $n = -2$ | $2.67 \pm 0.06$ | $0.26 \pm 0.05$ | 0.9[15] |
| $n = -1$ | $2.66 \pm 0.05$ | $0.21 \pm 0.03$ | 2.2[16] |
| $n = 0$ | $2.70 \pm 0.05$ | $0.10 \pm 0.03$ | 1.7[18] |

Table 4. Four-Parameter Fits to Corrected Density Profiles

| MODEL | $\eta$ | $\eta + \gamma$ | $\chi^2/\nu\ [\nu]$ |
|---|---|---|---|
| $\Omega_o = 1.0$ | | | |
| $n = -2$ | 1.66± 0.69 | 2.28 ± 1.01 | 0.7[15] |
| $n = -1$ | 2.21± 0.23 | 2.34 ± 0.35 | 1.5[15] |
| $n = 0$ | 2.34± 0.27 | 2.46 ± 0.42 | 0.8[14] |
| $\Omega_o = 0.2$ | | | |
| $n = -2$ | 1.94± 0.11 | 2.62 ± 0.17 | 3.0[18] |
| $n = -1$ | 2.29± 0.07 | 2.81 ± 0.12 | 1.9[20] |
| $n = 0$ | 2.38± 0.06 | 3.00 ± 0.25 | 1.2[19] |
| $\Omega_o = 0.1$ | | | |
| $n = -2$ | 2.10± 0.13 | 2.79 ± 0.20 | 1.7[20] |
| $n = -1$ | 2.12± 0.12 | 2.74 ± 0.15 | 1.0[25] |
| $n = 0$ | 2.43± 0.09 | 2.86 ± 0.18 | 1.3[22] |
| $\Omega_o = 0.2, \lambda_o = 0.8$ | | | |
| $n = -2$ | 1.89± 0.32 | 2.53 ± 0.49 | 1.1[14] |
| $n = -1$ | 2.01± 0.23 | 2.52 ± 0.36 | 2.8[15] |
| $n = 0$ | 2.39± 0.11 | 3.14 ± 0.18 | 1.3[17] |

## Figure Captions

Fig. 1.— Initial power spectra calculated from the initial density fields. Gaussian random density fields were generated by selecting from power spectra $P(k) \sim k^n$ with $n = 0, -1, -2$. There is some sampling noise in each realization at low $k$.

Fig. 2.— Comparison of average profiles for the 30 most massive groups divided into three mass ranges for the $\Omega = 1.0, n = -1$ run. Ten clusters are included in each average,m ranked 1–10, 11–20 and 21–30. For this run, the mass ranges are (in terms of the number of particles) 6749 − 1145, 1105 − 725, and 715 − 525.

Fig. 3.— (a) Average profiles of the 35 most massive clusters in each model. Also plotted are power law fits to the region $100 < \rho/\bar{\rho} < 3000$, with points at $r > 4\epsilon$ included. One–sigma error bars are given. For bins in which there was only one point, the error bars are set to half the value of the density. Residuals to the fits are shown in (b), where the horizontal axis has been expanded relative to (a) to show the region of the fit.

Fig. 4.— Logarithmic slopes $\alpha$ of the power law fits, as a function of the initial spectral index $n$. Lines connect the points for each cosmological model as indicated.

Fig. 5.— Mean radial velocity profiles, normalized to the circular velocity of each cluster at $r_\delta$. The radial velocity due to pure Hubble flow for each model is indicated by a dotted line.

Fig. 6.— Normalized velocity dispersion profiles for the runs. The radial dispersion is calculated with respect to the mean at each radius: $\sigma^2 = \langle v_r^2 \rangle - \langle v_r \rangle^2 + 2\langle v_t^2 \rangle$.

Fig. 7.— Velocity anisotropy profiles for the runs.

Fig. 8.— The effect of force softening on the force $F_s/F$ arising from a spherical mass distribution. The ratio of softened to unsoftened force is given for a point mass and for objects with $\rho \sim r^{-\alpha}$, where $\alpha = 2$ and 3. A density cutoff is imposed to make the object more physical and to allow the calculation of forces when $\alpha \geq 3$ (see text).

Fig. 9.— Comparison of the density profiles in our standard runs (solid lines) with those from simulations in which $\epsilon$ is increased by a factor of two (dashed lines). For each model, there turnover at small radii in runs with larger $\epsilon$ is more apparent. Data points indicate results of applying our correction procedure to estimate the small–$\epsilon$ results from the large-$\epsilon$ profiles.

Fig. 10.— Density profiles corrected for the effects of force softening. Two fits are also included: power law fits to the region $100 < \rho/\bar{\rho} < 3000$, and quadratic fits to $\rho/\bar{\rho} > 100$.

Fig. 11.— Circular velocity profiles $\sqrt{GM/r}$ derived from corrected (solid) and uncorrected (dotted) cluster mass profiles.

Fig. 12.— Circular velocity data from the $\Omega = 1$, $n = -1$ simulations of Quinn *et al.* (1986) (solid lines) and Efstathiou *et al.* (1988) (dashed lines) plotted against radius expressed in terms of an effective gravitational softening length (see text). The decline in the circular velocities at large radii is consistent with the density profile slope $\alpha = -2.33$ (solid, bold line).